\documentclass[preprint,authoryear,12pt]{elsarticle}

\usepackage{color}
\usepackage{rotating}
\usepackage{hyperref}
\usepackage{graphicx}
\usepackage{amsmath}
\usepackage{amsfonts}
\usepackage{amssymb}
\usepackage{rotating}
\usepackage[english]{babel}

\newcommand*\sun{\ensuremath{\odot}}

\def\la{\mathrel{\mathchoice {\vcenter{\offinterlineskip\halign{\hfil
$\displaystyle##$\hfil\cr<\cr\sim\cr}}}
{\vcenter{\offinterlineskip\halign{\hfil$\textstyle##$\hfil\cr
<\cr\sim\cr}}}
{\vcenter{\offinterlineskip\halign{\hfil$\scriptstyle##$\hfil\cr
<\cr\sim\cr}}}
{\vcenter{\offinterlineskip\halign{\hfil$\scriptscriptstyle##$\hfil\cr
<\cr\sim\cr}}}}}
\def\ga{\mathrel{\mathchoice {\vcenter{\offinterlineskip\halign{\hfil
$\displaystyle##$\hfil\cr>\cr\sim\cr}}}
{\vcenter{\offinterlineskip\halign{\hfil$\textstyle##$\hfil\cr
>\cr\sim\cr}}}
{\vcenter{\offinterlineskip\halign{\hfil$\scriptstyle##$\hfil\cr
>\cr\sim\cr}}}
{\vcenter{\offinterlineskip\halign{\hfil$\scriptscriptstyle##$\hfil\cr
>\cr\sim\cr}}}}}

\def \K{~\rm{K}}

\def \AU{~\rm{AU}}

\def \yr{~\rm{yr}}

\def \kpc{~\rm{kpc}}

\def\aj{AJ~}
\def\apj{ApJ~}
\def\aap{A\&A~}
\def\apjl{ApJ~}
\def\apjs{ApJS~}

\def\araa{ARA\&A~}

\def\mnras{MNRAS~}

\journal{New Astronomy}

\begin{document}

\begin{frontmatter}

\title{Is  the   central  binary   system  of  the   planetary  nebula
  Henize~2--428 a Type Ia supernova progenitor?}

\author[1,2]{Enrique Garc\'\i a--Berro}
\ead{enrique.garcia-berro@upc.edu}
\author[3]{Noam Soker}
\ead{soker@physics.technion.ac.il}
\author[4]{ Leandro G. Althaus}
\ead{althaus@fcaglp.fcaglp.unlp.edu.ar}
\author[5,2]{Ignasi Ribas}
\ead{iribas@ice.csic.es}
\author[5,2]{Juan C. Morales}
\ead{morales@ieec.uab.es}
\address[1]{Departament de F\'\i sica Aplicada,
           Universitat Polit\`ecnica de Catalunya,
           c/Esteve Terrades 5,
           08860 Castelldefels, Spain}
\address[2]{Institut d'Estudis Espacials de Catalunya,
           c/Gran Capit\`a 2--4, Edif. Nexus 104,
           08034 Barcelona, Spain}
\address[3]{Department of Physics,
           Technion -- Israel Institute of Technology,
           Haifa 32000,
           Israel}
\address[4]{Facultad de Ciencias Astron\'omicas y Geof\'\i sicas,
           Universidad Nacional de La Plata--IALP-CONICET,
           Paseo del Bosque s/n, \\
           1900 La Plata,
           Argentina}
\address[5]{Institut de Ci\`encies de l'Espai--CSIC,
            Campus UAB, Facultat de Ci\`encies, 
            Torre C5 -- parell -- 2a planta
            08193 Bellaterra, 
            Spain}

\begin{abstract}
We critically discuss the recent  observations of the binary system at
the center of the bipolar planetary nebula Henize~2--428. We find that
the proposed explanation  of two equal-mass degenerate  objects with a
total mass larger than the Chandrasekhar limiting mass that supposedly
will merge in less than a Hubble time, possibly leading to a SN~Ia, is
controversial.   This hypothesis  relies  on the  assumption that  the
variability of the  He~{\sc II}~5412~\AA~ spectral line is  due to two
absorption components.  Instead,  we propose that it  can be accounted
for by a broad absorption line from the central system on top of which
there is a narrow emission line  from the nebula.  This prompted us to
study if  the binary  system can be  made of a  degenerate star  and a
low-mass  main sequence  companion, or  of two  degenerate objects  of
smaller mass.   We find that  although both scenarios can  account for
the existence  of two symmetric broad  minima in the light  curve, the
second one  agrees better with  observations.  We thus argue  that the
claim  that Henize~2--428  provides observational  evidence supporting
the double-degenerate scenario for SN~Ia is premature.
\end{abstract}


\end{frontmatter}

\section{Introduction}
\label{sec:introduction}

Thermonuclear, or Type  Ia supernovae (SNe~Ia), are the  result of the
explosion  of carbon-oxygen  white dwarfs.   Despite their  well known
observed properties, the nature of the progenitor systems that produce
a SNe~Ia event has not been hitherto elucidated, and several scenarios
have been proposed,  none of which gives a satisfactory  answer to all
the abundant observational material.   The scenarios can be classified
into six  categories --- see, for  instance, \cite{TsebrenkoSoker2015}
for   a   recent   discussion   of   some   of   the   channels,   and
\cite{WangHan2012}  and \cite{Maozetal2014}  for  extended reviews  of
some of these scenarios.

As there is  no consensus on which are the  SN~Ia progenitor(s), it is
crucial to  refer to all  scenarios (or categories of  scenarios) when
confronting  them  with observations.  We  list  them in  alphabetical
order,  and cite  only  a few  references for  each  scenario: a)  The
core-degenerate   (CD)   scenario  \citep{Livio2003,   KashiSoker2011,
Sokeretal2013},   b)  The   double-degenerate  (DD)   scenario  (e.g.,
\citealt{Webbink1984,  Iben1984}),  c)  The  double-detonation  (DDet)
mechanism (e.g.,  \citealt{Woosley1994, Livne1995,  Shenetal2013}.  d)
The  single-degenerate   (SD)  scenario   (e.g.,  \citealt{Whelan1973,
Nomoto1982, Han2004}),  e) The  recently proposed  singly-evolved star
(SES)  scenario \citep{Chiosietal2015},  and  f)  The WD-WD  collision
(WWC)    scenario   (e.g.,    \citealt{Raskinetal2009,   Thompson2011,
Kushniretal2013, Aznar2013}).

Since all these scenarios involve white dwarfs, all progenitors evolve
through one or two planetary  nebula (PN) phases.  Accordingly, one of
the pieces of evidence that would help in constraining SN~Ia scenarios
is to  study PNe.   Furthermore, in  some cases  SN~Ia have  been even
claimed   to    take   place    inside   planetary    nebulae   (e.g.,
\citealt{DickelJones1985, TsebrenkoSoker2013,  TsebrenkoSoker2015}), a
process termed SNIP.

In a recent paper  \cite{SantanderGarciaetal2015} analyzed the central
binary    system    of     the    planetary    nebula    Henize~2--428
\citep{Rodriguezetal2001,                    SantanderGarciaetal2011}.
\cite{SantanderGarciaetal2015} found  that the light curve  of this PN
shows two nearly identical  broad minima, indicating significant tidal
distortion of  the components of binary  system, and that there  is an
absorption line of He~{\sc II}~5412~\AA~ that varies with time.  Given
that the two minima of the light curve are practically identical, they
assumed  that they  are  caused by  a binary  system  composed of  two
equal-mass stars of the same  type, and found the temperature, radius,
and luminosity, of the two stars  to be almost identical. They further
argued that  most likely these  are two degenerate stars,  i.e., white
dwarfs or cores of post-asymptotic  giant branch (AGB) stars, on their
way to become CO white dwarfs.  As  the combined mass in this model is
$1.76\, M_{\sun}$, \cite{SantanderGarciaetal2015}  further argued that
these two  stars will merge  to form  a SN~Ia in  the frame of  the DD
scenario.

Here     we     critically      discuss     the     explanation     of
\cite{SantanderGarciaetal2015}.       As       we      explain      in
Sect.~\ref{sec:considerations}  we  find  the  interpretation  of  the
observations of \cite{SantanderGarciaetal2015} to be plausible, albeit
other  possibilities  are  conceivable.  In  Sect.~\ref{sec:model}  we
relax  the assumptions  of these  authors and  we propose  alternative
models of  the binary system.  The  first of these models  consists of
binary system  in which  only one  of the  components is  a degenerate
star, while the secondary star is normal non-evolved star.  The second
of the models involves two  non-identical degenerate stars, but with a
combined mass smaller  than the Chandrasekhar limiting  mass.  A short
summary is given in Sect.~\ref{sec:summary}.

\section{Preliminary considerations}
\label{sec:considerations}

\subsection{A binary system made of two identical stars?}
\label{subsec:equal}

\cite{SantanderGarciaetal2015} argue  for a binary system  composed of
two stars having  the same mass, $0.88 \pm 0.13\,  M_{\sun}$, the same
luminosity, $\approx 420 \, L_{\sun}$ at a distance of $1.4 \kpc$, and
the same radius,  $0.68 \pm 0.04 \, R_{\sun}$.  This  implies that the
two  stars are  at the  same evolutionary  stage.  However,  any small
difference in the main  sequence mass will turn to a  large one on the
asymptotic giant  branch (AGB). An AGB  star having a core  of $0.88\,
M_{\sun}$  burns  hydrogen at  a  rate  of  $\sim 2  \times  10^{-7}\,
M_{\sun} \yr^{-1}$ (e.g.,  \citealt{Paczynski1970}).  For a difference
in mass  between the two cores  $<0.02 \, M_{\sun}$ the  difference of
evolutionary  times between  the post-AGB  stars should  be $\la  10^5
\yr$.  This requires a mass difference on the main sequence of $\Delta
M/M \la 10^{-3}$, depending on the initial mass of the stars.

It could  be argued that  there are  other binary systems  with almost
identical  components, known  as twin  binaries \citep{LucyRicco1979}.
Specifically,   \cite{PinsonneaultStanek2006}   studied  21   detached
eclipsing binaries in  the Small Magellanic Cloud and  found that 50\%
of  detached  binaries  have  companions  with  very  similar  masses.
However, \cite{Lucy2006} and  \cite{CantrellDougan2014} concluded that
there is  a strong  observational bias that  affects spectroscopically
selected binary  stars, and  that the  apparent overabundance  of twin
binaries does not reflect their true population.  In summary, the case
for a twin binary is possible,  but unlikely, hence motivates us for a
careful reexamination of such a claim.

\subsection{Stellar properties}
\label{subsec:stellar}

As mentioned, in the  model proposed by \cite{SantanderGarciaetal2015}
each star is a post-AGB star with  a mass of $0.88 \, M_{\sun}$.  When
a post-AGB  of that mass  fades to a luminosity  of $ \approx  10^3 \,
L_{\sun}$  its radius  is  already $\simeq  0.02  \, R_{\sun}$  (e.g.,
\citealt{BloeckerSchoenberner1991}).   This radius  is about  30 times
smaller than  the radius suggested  by \cite{SantanderGarciaetal2015}.
This poses a serious problem to their model.

In the first of our models  we investigate a case where the luminosity
of the  system is  due to  just one  star, and  the luminosity  of the
companion is negligible --- see below for more details.  At a distance
of  $D=1.4  \kpc$  as deduced  by  \cite{SantanderGarciaetal2015}  the
luminosity is  $\approx 850 \,  L_{\sun}$.  This  can be a  star whose
evolution was truncated on the upper  red giant branch (RGB), when its
core mass was only $M_1 \approx 0.45 \, M_{\sun}$, or on the lower AGB
when its core mass was $\approx  0.5 \, M_{\sun}$.  On the other hand,
if  the  distance is  larger,  say  $D=1.8  \kpc$, the  luminosity  is
$\approx  1.4\times  10^3\,  L_{\sun}$.   This can  be  a  star  whose
evolution was truncated on the lower  AGB, when its core mass was only
$M_1  \approx  0.52-0.55  \,  M_{\sun}$. In  our  proposed  model  the
companion that terminated the RGB or  the AGB evolution of the primary
component is  a main sequence star  of $\sim 0.3-0.5 \,  M_{\sun}$. In
the second of our models we assume that indeed both stars are post-AGB
stars but  we allow the  stars to have different  physical parameters,
namely different masses, effective temperatures and luminosities.

A  note   is  in   place  here  on   the  distance   to  Henize~2-428.
\cite{SantanderGarciaetal2015}  provided  a   rough  estimate  of  the
distance  of $1.4  \pm  0.4  \kpc$ based  on  the dereddened  apparent
magnitudes of Henize~2--428. \cite{Maciel1984}  obtained a distance of
$1.7 \kpc$, \cite{CahnKaler}  derived a distance of  $2.7 \kpc$, while
the  most recent  determination  of \cite{Frew},  using the  H$\alpha$
surface brightness--radius relation is also $2.7 \kpc$. Based on these
values we will scale our  expressions with two distances, $D=1.4 \kpc$
and     $D=1.8     \kpc$,     as      the     value     adopted     by
\cite{SantanderGarciaetal2015}  was obtained  from  their  fit to  the
properties of the binary system, which is questioned here.

\subsection{Light curves and spectrum}
\label{subsec:observations}

The arguments  of \cite{SantanderGarciaetal2015} for their  claim of a
binary system of  equal-mass stars at the same  evolutionary stage are
the nearly identical  minima in the light curve, and  the line profile
of the He~{\sc II}~5412~\AA~ spectral  feature --- see their figures 2
and  3.  The  nearly identical  minima of  the light  curve have  been
suggested to be indicative that both members of the binary system have
very  similar  masses.   Additionally,  \cite{SantanderGarciaetal2015}
found that the He~{\sc  II}~5412~\AA~spectral line of Henize~2--428 is
variable.  They  attributed the  variability of  this line  to Doppler
shifts of two equal-mass stars,  and then used two Gaussian absorption
profiles  to  model  the  variation.   Consequently,  in  their  joint
analysis of  the light curve  and the  spectrum they {\sl  forced} the
mass ratio  $q=M_2/M_1$ of  the binary system  to be  1.  Furthermore,
\cite{SantanderGarciaetal2015} {\sl did not model the spectra} of both
components of the  binary system, since they were not  able to measure
surface  gravities for  each  one of  the  individual binary  members.
Finally,  they  {\sl  assumed}  that   both  stars  are  at  the  same
evolutionary stage.   All of these  assumptions are critical  in their
analysis.

In particular,  it must  be stressed  that even if  the mass  ratio is
close to 1,  the nature of the  stars can be very  different, and that
the  lack  of determinations  of  surface  gravities leaves  room  for
alternative explanations.   In particular, a  close look at  figures 2
and 3 of \cite{SantanderGarciaetal2015}  suggests that the spectrum of
Henize~2--428 can be  explained by assuming that there  is an emission
line  on  top of  the  absorption  profile.   We therefore  examine  a
possible  alternative interpretation  where  the line  profile is  the
result of a wide absorption line  belonging to the primary star, and a
narrow emission line coming from  the compact dense nebula reported by
\cite{Rodriguezetal2001}, or which originates  even much closer to the
star  from the  wind  itself.  In  this  alternative explanation  both
emission and absorption lines change  with orbital phase.  This is not
unusual.  Many  central stars  of planetary  nebulae show  He~{\sc II}
absorption  lines (e.g.,  \citealt{WeidmannGamen2011}).  The  emission
line is seen in some nebulae,  e.g., in Abell~48 whose central star is
a WN star  \citep{Todtetal2013}.  Most interestingly, in  the study of
\cite{WeidmannGamen2011}  there  are  several  PNe that  show  a  wide
He~{\sc II}~5412~\AA~ absorption line with  a weak emission feature in
the  center  of the  wide  absorption  line.   This forms  a  spectral
structure  similar  to that  of  Henize~2--428.   The most  noticeable
examples of this are the PNe  He~2--105, He~2--434, and to some degree
SP~3 and PC~12. All these central stars  are O stars.  To these we add
the  PN  Pa~5,  for  which  the central  star  shows  a  wide  He~{\sc
II}~5412~\AA~ absorption line  and the resolved nebula  shows a narrow
He~{\sc II}~5412~\AA~  emission line  \citep{Garciaetal2014}.  Another
relevant case is the central star  of the Eskimo planetary nebula (NGC
2392). The He~{\sc II}~5412~\AA line shows a structure similar to that
of Henize~2--428 \citep{PrinjaUrbaneja2014}.  This structure varies on
time scales down to about an hour. \cite{PrinjaUrbaneja2014} attribute
the variation in the line to  a variable distribution of clumps in the
wind, variations  in the  velocity field, and/or  the mass  loss rate,
rather to a binary star.

It could  be argued  that if  our interpretation  is correct,  and the
He~{\sc   II}~5412~\AA~   emission   line  is   nebular,   a   He~{\sc
II}~4686~\AA~emission  line  should  be   also  observed,  but  it  is
not. However,  other PNe,  like He~2--105,  PC~12, He~2--434  and SP~3
have spectra with weak  He~{\sc II}~5412~\AA~bumps (which we interpret
as an emission line) in the center of a broad absorption line, whereas
the      He~{\sc      II}~4686~\AA~line     is      in      absorption
\citep{WeidmannGamen2011}. On  the other hand,  in the PN  M~1--14 the
He~{\sc   II}~5412~\AA~line  is   in   absorption   but  the   He~{\sc
II}~4686~\AA~occurs  in   emission  \citep{WeidmannGamen2011}.   These
cases show that there is  an interplay between emission and absorption
in the He~{\sc II} lines. This suggests that the emission region, like
the absorption one, is located very close  to the star, but not at the
same  place.  In  all these  cases  both the  absorption and  emission
features vary with orbital phase.  In conclusion, based on the current
observations the  existence of a He~{\sc  II} emission line on  top of
the  broad absorption  feature cannot  be discarded,  and needs  to be
considered.

Furthermore, \cite{SantanderGarciaetal2015} established an upper limit
on the  effective temperature of the  members of the binary  system of
$\approx 40,000 \K$,  based on the absence of  a He~{\sc II}~5412~\AA~
emission  line,  but there  are  many  PNe  whose central  stars  have
effective temperatures larger than $40,000 \K$ and do not have He~{\sc
II} emission lines.  Since we assume  that the emission line is indeed
present     in     the     spectrum,    the     upper     limit     of
\cite{SantanderGarciaetal2015}  to  the  effective temperature  is  no
longer  valid, and  the  effective temperature  could be  sufficiently
large to allow emission.  Consequently, in the following we will scale
quantities with an effective temperature $T_{\rm eff}=45,000 \K$.

\section{Possible binary models}
\label{sec:model}

According  to  all the  considerations  put  forward in  the  previous
section,  here we  explore  two  models in  which  the  mass ratio  is
$\approx 1$, but  with different characteristics.  The  first of these
models involves  a semi-detached  system in which  the members  of the
binary system  are not at the  same evolutionary stage, while  for the
second  model  we adopt  an  overcontact  binary  system made  of  two
post-AGB stars,  as \cite{SantanderGarciaetal2015} did, but  with less
restrictive assumptions.

\subsection{A semi-detached binary system}
\label{subsec:present}

We first examine a binary model where the secondary is a main sequence
star of mass $M_2 \simeq 0.3-0.5 M_{\sun}$.  If we take the primary to
be a post-RGB star of mass $M_1 \simeq 0.45 M_{\sun}$ or a post-AGB of
mass $M_1  \simeq 0.55 M_{\sun}$ (e.g.,  \citealt{Bloecker1993}), then
these two cases span a mass ratio of $q=M_2/M_1\simeq 0.6-1.0$.  Using
the  expression   for  the   Roche  lobe   radius  $r_{\rm   L}$  from
\cite{Eggleton1983}, we find for  the primary star $r_{\rm L1}/a=0.42$
and $0.38$ for  $q=M_2/M_1=0.6$ and $1.0$, respectively,  where $a$ is
the  orbital  separation.  For  the  secondary  star we  find  $r_{\rm
L2}/a=0.34$ and $0.38$ for $q=M_2/M_1=0.6$ and $1$, respectively.  For
an orbital period of $P=4.2$~h  the orbital separation (for a circular
orbit)  is $a=  1.27 ({M}/{0.9  \,M_{\sun}})^{1/3} R_{\sun}$,  and the
Roche lobe of the primary star is
\begin{equation}
 r_{\rm L1}= 0.51 \left( \frac{M}{0.9 \, M_{\sun}} \right)^{1/3}
 \left( \frac{r_{\rm L}/a}{0.4} \right)\, R_{\sun},
 \label{eq:Roche}
\end{equation}
where  $M=M_1+M_2$  is  the  total  binary  mass.   From  the  primary
luminosity, $L_1= 845 (D/1.4\, {\rm kpc})^{2}\, L_{\sun}$, and for the
effective temperature assumed here, the primary radius is
\begin{equation}
 R_1= 0.48 \left( \frac{D}{1.4 \kpc} \right)
 \left( \frac{T_1}{4.5 \times 10^4 \K} \right)^{-2}\,
 R_{\sun}.
 \label{eq:radius}
\end{equation}
For a distance of $D=1.8 \kpc$  the primary radius with that effective
temperature  is  $R_1=0.62  \,  R_{\sun}$, but  taking  a  temperature
$T_1=50,000\K$  will  make  the  primary just  filling  in  its  Roche
lobe. In  our model  the primary  is close to  filling its  Roche lobe
(assuming synchronization). For the secondary the Roche lobe radius is
in  the  range $r_{\rm  L2}  \simeq  0.42-0.5\, R_{\sun}$.  Thus,  the
secondary also touches its Roche lobe.

We  consider  two  possible  evolutionary scenarios.   We  prefer  the
post-RGB one, but cannot rule out  the post-AGB one. We first describe
the relevant  evolutionary tracks on the  Hertzsprung-Russell diagram,
and then turn to discuss the  two possible scenarios. The last step of
our analysis  consists of  computing the light  curve of  the proposed
binary system.

\subsubsection{Evolutionary tracks}
\label{subsubsec:tracks}

\begin{figure}
   \begin{center}
   \includegraphics[width=0.7\columnwidth,clip=true]{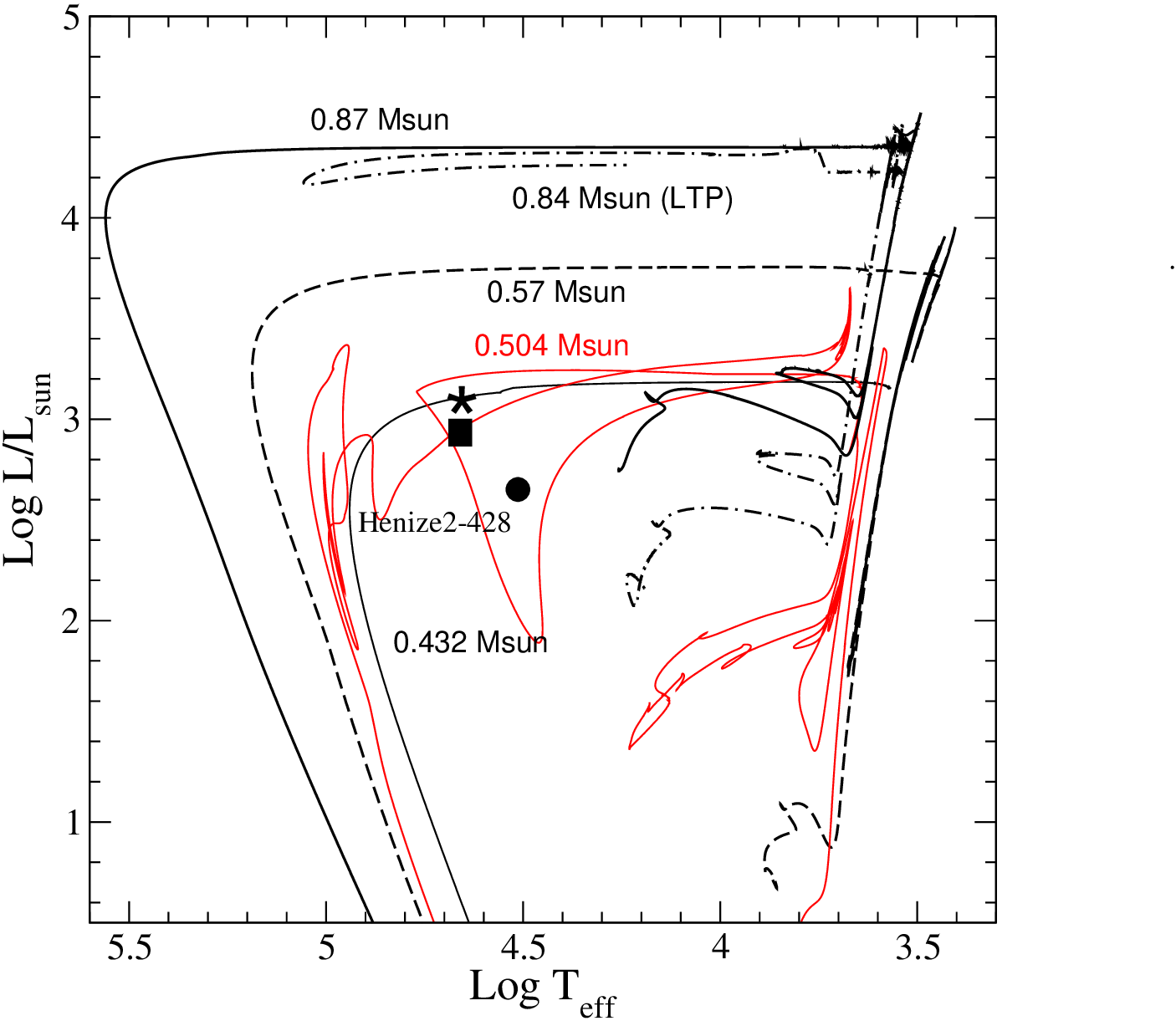}
   \\
   \includegraphics[width=0.7\columnwidth,clip=true]{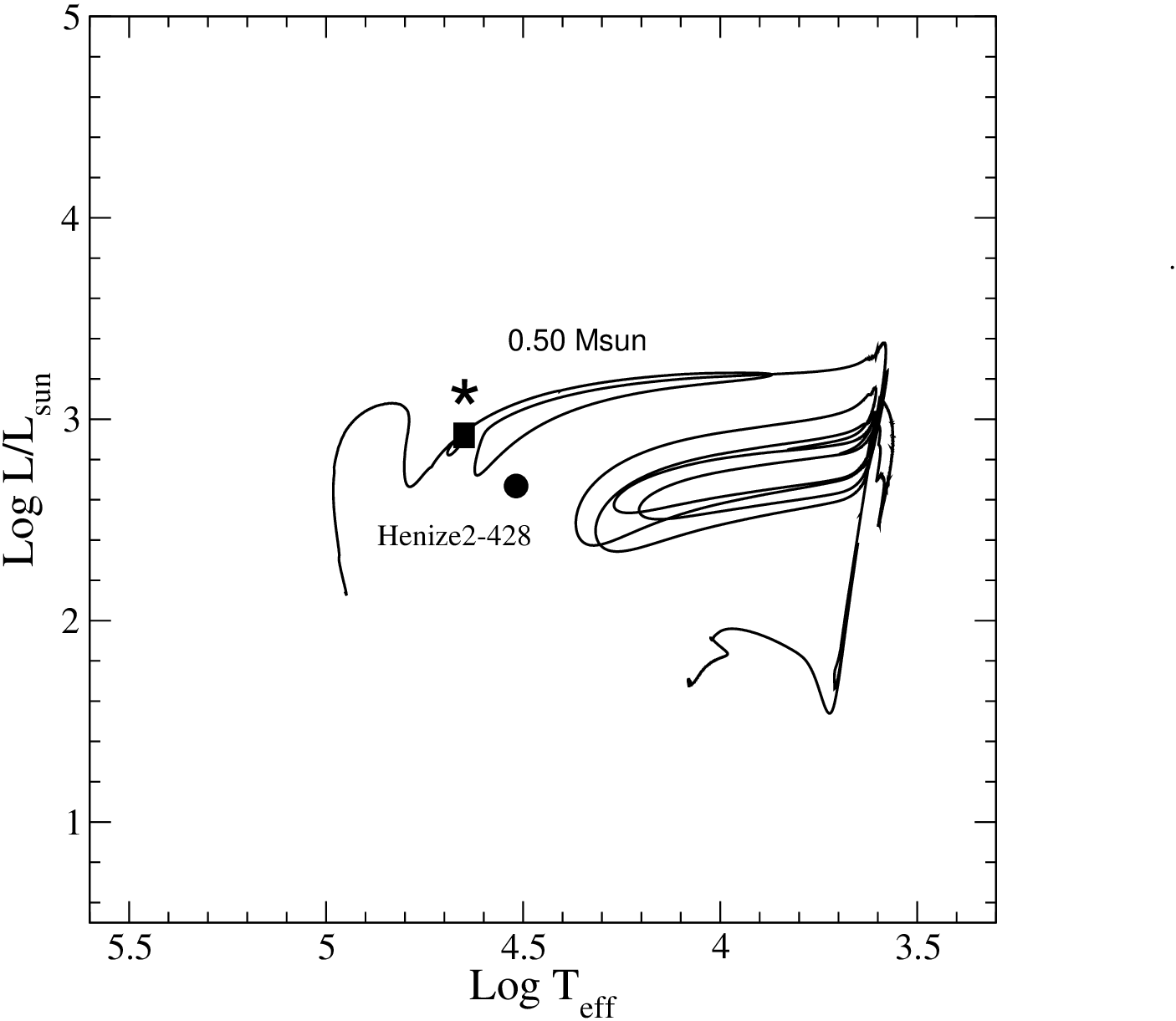}
   \caption{Relevant    evolutionary    tracks     (see    text    for
     details). Marked  are the location  of each of the  two identical
     stars   of    Henize~2--428   according    to   the    model   of
     Santander-Garc\'\i  a  et  al.  (2015), filled  circle,  and  the
     central  star  in our  proposed  scenario  when the  distance  is
     $D=1.4$~kpc (filled square) and when $D=1.8$~kpc (asterisk).}
    \label{fig:HR}
   \end{center}
\end{figure}

In  the top  panel  of  Fig.  \ref{fig:HR}  we  present some  relevant
evolutionary tracks on the  Hertzsprung-Russell diagram, together with
the approximate location  of the central star of  Henize~2--428 at two
distances, $D=1.4  \kpc$ (square) and  $1.8 \kpc$ (asterisk),  and the
position  reported  by  \cite{SantanderGarciaetal2015},  circle.   The
sequences of masses $M_1=0.57$ and  $0.87 \, M_{\sun}$ were taken from
\cite{renedo}, and  are the result  of the full evolution  of zero-age
main  sequence (ZAMS)  stars  with masses  $M_{{\rm ZAMS}_1}=1.5$  and
$5.0\, M_{\sun}$,  respectively, of Solar metallicity.   The evolution
is followed  through all the  relevant stages, including  the hydrogen
and helium core  burning phases, the thermally-pulsing  AGB phase, and
the post-AGB  evolution to  the white  dwarf stage.   The $M_1=0.504\,
M_{\sun}$ sequence was also  taken from \cite{renedo}, and corresponds
to  a progenitor  of  mass $M_{{\rm  ZAMS}_1}=0.85  \, M_{\sun}$  with
$Z=0.001$.  In  addition,  the  post-RGB  evolution  of  a  $0.432  \,
M_{\sun}$ helium-core low-mass white dwarf is included.  This sequence
is the  result of the non-conservative  binary evolution of a  star of
mass $M_{{\rm  ZAMS}_1}=1.0\, M_{\sun}$  that abandons the  RGB before
the  onset  of  core  helium  burning  \citep{leandro}.  Finally,  the
evolutionary track of  a $M_1=0.84 \, M_{\sun}$ post-AGB  remnant of a
$M_{{\rm   ZAMS}_1}=3.0\,   M_{\sun}$  progenitor   with   metallicity
$Z=0.001$  that  experiences  a  late thermal  pulse  (LTP)  when  its
effective temperature was $T_{\rm eff}=10,000$~K is shown as well.  As
a result of the LTP, the post-AGB remnant experiences a fast evolution
to the  blue.  After  reaching a  maximum effective  temperature, this
remnant evolves back to the domain of giant stars.

Note that the luminosity inferred for Henize~2--428, for the effective
temperatures  used by  \cite{SantanderGarciaetal2015}  and  by us,  is
substantially  smaller than  that predicted  by post-AGB  evolutionary
models for masses of $M_{{\rm ZAMS}_1} \ga 0.55 M_{\sun}$. The claimed
masses for the two central stars of Henize~2--428 do not fit the other
claimed   properties  of   the  stars   in  the   model  proposed   by
\cite{SantanderGarciaetal2015}.  However,  the $M_1=0.504\,  M_{\sun}$
post-AGB  star presented  here  (upper panel)  experiences  an LTP  at
$T_{\rm eff}=48,000 \K$, which brings  the remnant back rapidly to the
red  giant domain  and  finally  to the  white  dwarf  stage. Such  an
evolutionary track covers  the general region of  Henize~2--428 on the
Hertzsprung-Russell diagram.

Finally, the bottom panel  of Fig.~\ref{fig:HR} displays the evolution
of a star  with $M_{{\rm ZAMS}_1}=2.5\, M_{\sun}$,  and $Z=0.01$.  The
evolution  of this  star  was truncated  on the  lower  AGB, when  its
luminosity is  $L=1.4\times 10^3\, L_{\sun}$, the  observed luminosity
of the  binary system for the  distance we adopt.  At  this point mass
was removed to mimic a common-envelope  episode. The final mass of the
remnant  after   this  intense  episode   of  mass  loss   is  $0.50\,
M_{\sun}$. The  evolution in  the Hertzsprung-Russell diagram  of this
sequence  shows  several  blue  loops, which  are  due  to  successive
ignitions of  the hydrogen shell. As  can be seen, this  track is also
able  to reproduce  the observed  position of  Henize~2--428, for  our
adopted distance.

In conclusion,  post-RGB helium-core  remnants with stellar  masses of
$M_1  \approx0.45 \,M_{\sun}$  and low-mass  cores of  AGB stars  that
truncate  the  lower  AGB  with masses  of  $M_1  \approx  0.50-0.55\,
M_{\sun}$ can account for the central star of Henize~2--428.

\subsubsection{Post-RGB evolution}
\label{subsubsec:postRGB}

\cite{Guerreroetal2000}   studied  15   bipolar  PNe,   and  only   in
Henize~2--428 they detect no H$_2$ emission, and it was the only PN in
their sample in which a bright  central star was found. It seems there
is  something strange  in  the evolution  of this  PN,  which here  is
attributed    to     the    system    being    a     post-RGB    star.
\cite{Rodriguezetal2001}  found the  low abundances  of most  elements
they study  (relative to hydrogen)  of Henize~2--428 to be  similar to
those found  for PNe belonging  to the Galactic halo.   They suggested
that this points  to a low-mass progenitor. It is  quite possible that
the central  star of  Henize~2--428 is  a post-RGB  star orbited  by a
low-mass main sequence star. The  post-RGB scenario is compatible with
the low nebular mass, as the  initial stellar mass in this scenario is
$M_{{\rm ZAMS}_1} \approx 1\, M_{\sun}$, and the post-RGB mass is $M_1
\approx 0.45\, M_{\sun}$.

\subsubsection{Post-AGB evolution}
 \label{subsubsec:postAGB}

In this  proposed alternative  scenario the  evolution of  the primary
star is  truncated on  the lower  AGB when  its luminosity  is between
$\approx  10^3$ and  $1.5\times  10^3\, L_{\sun}$  and  its radius  is
$\approx 150\, R_{\sun}$,  assuming in this case a  distance of $D=1.8
\kpc$.   The  core  mass  is  $\approx  0.50-0.55\,  M_{\sun}$  (e.g.,
\citealt{Bloecker1993}; Fig.  \ref{fig:HR}  here).  The companion will
spiral-in due to tidal forces when the primary radius is about quarter
of  the  orbital  separation \citep{Soker1996}.  The  initial  orbital
separation in this scenario is therefore $a_0 \approx 3 \AU$.

To avoid engulfment already on the  RGB phase of the primary star, the
primary size on the RGB must  be $<10^2 \, R_{\sun}$, which limits the
initial   mass  of   the  primary   star  to   be  $2.3   \la  M_{{\rm
ZAMS}_1}/M_{\sun} \la 6.0$ \citep{IbenTutukov1985}. Therefore, in this
scenario the  AGB star is more  massive, and a more  massive nebula is
expected.

\subsubsection{Light curve}
 \label{subsubsec:LC}

\begin{figure}
   \begin{center}
   \includegraphics[width=0.8\columnwidth,clip=true]{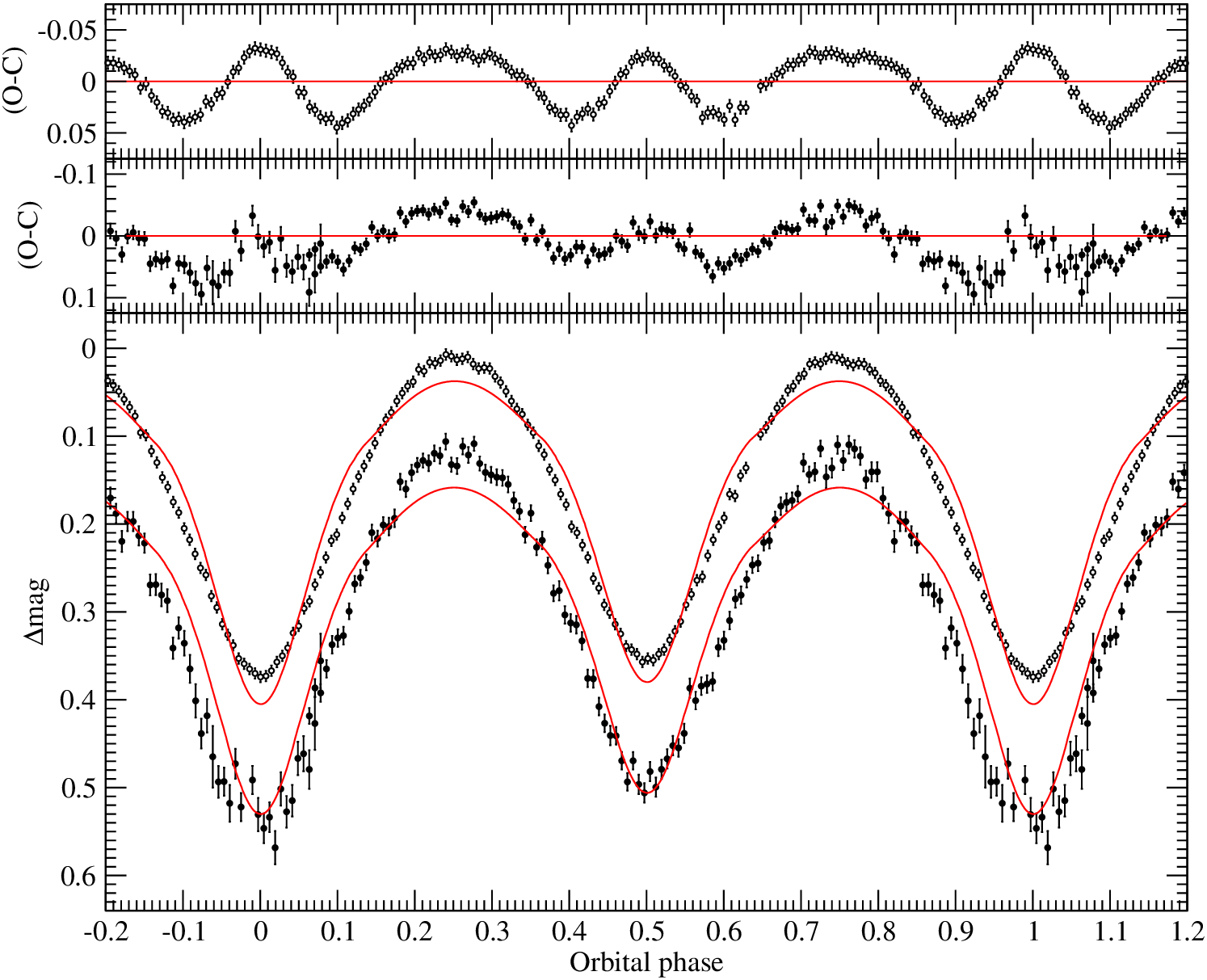}
   \caption{Light curve fit assuming a semidetached binary system made
     of  a degenerate  core and  a main  sequence companion,  with the
     primary component  filling the  Roche lobe.  B- and  I-band light
     curves  are  shown in  the  lower  panel  using filled  and  open
     circles, respectively. Red lines are the best-fit models for each
     band. Residuals are shown in the upper panels. Note the different
     scale for each band.}
   \label{fig:semidetached}
   \end{center}
\end{figure}

Now the natural question is whether  this model is consistent with the
observations of \cite{SantanderGarciaetal2015}.  Clearly, in our model
both  components of  the binary  system have  similar masses,  and the
stars are  very close.  Thus,  the minima of  the light curve  will be
broad, as a  consequence of tidal distortion of the  components of the
binary system.   This is in  accordance with the observed  light curve
obtained  by  \cite{SantanderGarciaetal2015}.   However,  there  is  a
fundamental    difference   between    our   model    and   that    of
\cite{SantanderGarciaetal2015}.  Due to  the large luminosity contrast
between the  evolved object and the  main sequence star, in  our model
the secondary  star can be easily  unnoticed in a first  analysis, and
the minima of the light curve are due to ellipsoidal variations of the
post-AGB star. However,  the light curve of this binary  system can be
easily confused with that of a  binary system in which both components
are post-AGB stars.

\begin{table}
\begin{center}
\begin{tabular}{lcc}
\hline
\hline
 & Primary & Secondary \\
\hline
$M$ ($M_{\sun}$)          & 0.495  & 0.396 \\
$R$ ($R_{\sun}$)          & 0.506  & 0.438 \\
\hline
$r$ pole                 & 0.3748 & 0.3226 \\
$r$ point                & 0.5230 & 0.4044 \\     
$r$ side                 & 0.3949 & 0.3406 \\
$r$ back                 & 0.4250 & 0.3669 \\
\hline
\hline
\end{tabular}
\end{center}
\caption{Absolute masses  and volumetric  radii derived from  the best
  fit to light curves assuming a semi-detached binary system. For each
  component, the  radius relative  to the semimajor  axis is  given in
  different directions  accounting for  deformation: towards  the pole
  ($r_{\rm  pole}$)  and  towards  the  companion  direction  ($r_{\rm
  point}$),  opposite  ($r_{\rm  back}$),  and  perpendicular  to  the
  orbital plane ($r_{\rm side}$).
  \label{tab:semidetached}}
\end{table}

To check  whether our model  is consistent  with the observed  data we
computed synthetic  light curves  for this model.  We used  the latest
version   of  the   Wilson-Devinney  code   \citep{WilsonDevinney1971,
Wilson1979}.    This  code   allows  to   fit  several   light  curves
simultaneously providing  parameters that are consistent  with all the
observed  data.  We  have  used  the B-  and  I-band  light curves  of
Henize~2--428  to  get the  orbital  and  physical parameters  of  the
system.  Specifically,  we obtained  the inclination with  respect the
visual plane ($i$), the pseudo-potentials ($\Omega_1$ and $\Omega_2$),
from  which  the  radii  of   the  components  can  be  computed,  the
temperature  of  the  secondary  component  ($T_{\rm  eff,  2}$),  the
luminosity ratio for each bandpass ($L_2$/$L_1$), and a phase shift to
account   for   period   inaccuracies.    Square-root   limb-darkening
coefficients were  interpolated in  the tables  of \cite{Claret2000a},
and  adjusted dynamically  according to  the current  temperatures and
surface  gravities of  the  stars at  each  iteration. The  reflection
albedos  were  fixed at  1.0,  appropriate  for stars  with  radiative
envelopes,  and  the gravity  darkening  exponents  were set  to  1.0,
following   \cite{Claret2000b}.   The   best   fit   is  obtained   by
differential corrections,  iterating until the internal  errors of the
parameters  are   smaller  than  the  corrections   applied  in  three
consecutive  iterations   and  repeating  this  process   five  times,
selecting as best fit that with the smaller residuals.

For  this  model we  fixed  the  orbital  separation  to $a  =  1.27\,
R_{\sun}$,  which corresponds  to  the observed  period  $P =  0.1758$
days. For the effective temperature  of the primary we adopted $T_{\rm
eff,1}=45,000$~K,   as    explained   in   Sect.~\ref{subsec:present}.
Finally,  the mass  ratio was  kept fixed  at $q=0.8$.   We fixed  the
primary   component    pseudo-potential   to   its    critical   value
($\Omega_1$=3.42 according to the mass ratio) to force a semi-detached
configuration and  we tested fits  with fixed values of  $\Omega_2$ in
steps of 0.1. This was done in order to prevent convergence towards an
overcontact system if $\Omega_2$ is  set free.  Nevertheless, the best
fit was found close to  the limit of overcontact with $\Omega_2=3.50$.
Figure~\ref{fig:semidetached}  and Table~\ref{tab:semidetached}  show,
respectively, the light curves in the  B- and I-band, and the relevant
physical parameters of the best-fit  solution.  The best fit model has
an  inclination $i=69.792^\circ$,  a phase  shift of  0.0008, and  the
effective   temperature    of   the   secondary   star    is   $T_{\rm
eff,2}=41,220$~K.   Finally,  the  ratio  of  the  luminosity  of  the
secondary star and  the primary component is 0.683 in  the B-band, and
0.679 in  the I-band.  Although  the residuals  of the fit  are rather
small (32.65~mmag  in the B-band,  and 23.51~mmag in the  I-band), the
top  panel   of  Fig.~\ref{fig:semidetached}  shows  that   there  are
systematic departures  of the computed  light curve from  the observed
one, which  can be also easily  noticed near the maxima  and minima of
the light  curve in the  bottom panel of  this figure. We  also tested
different mass ratios, from 0.6 to  1.0, but the light curve fits were
not significantly better.

Clearly, the  best-fit model is  at odds with our  initial hypothesis.
Most significantly,  the best-fit model  shows that the  luminosity of
the   secondary  star   is   comparable  to   that   of  the   primary
component. This motivates us to look for other possibilities.

\subsection{An overcontact binary system}
\label{subsec:overcontact}

\begin{figure}
   \begin{center}
   \includegraphics[width=0.8\columnwidth,clip=true]{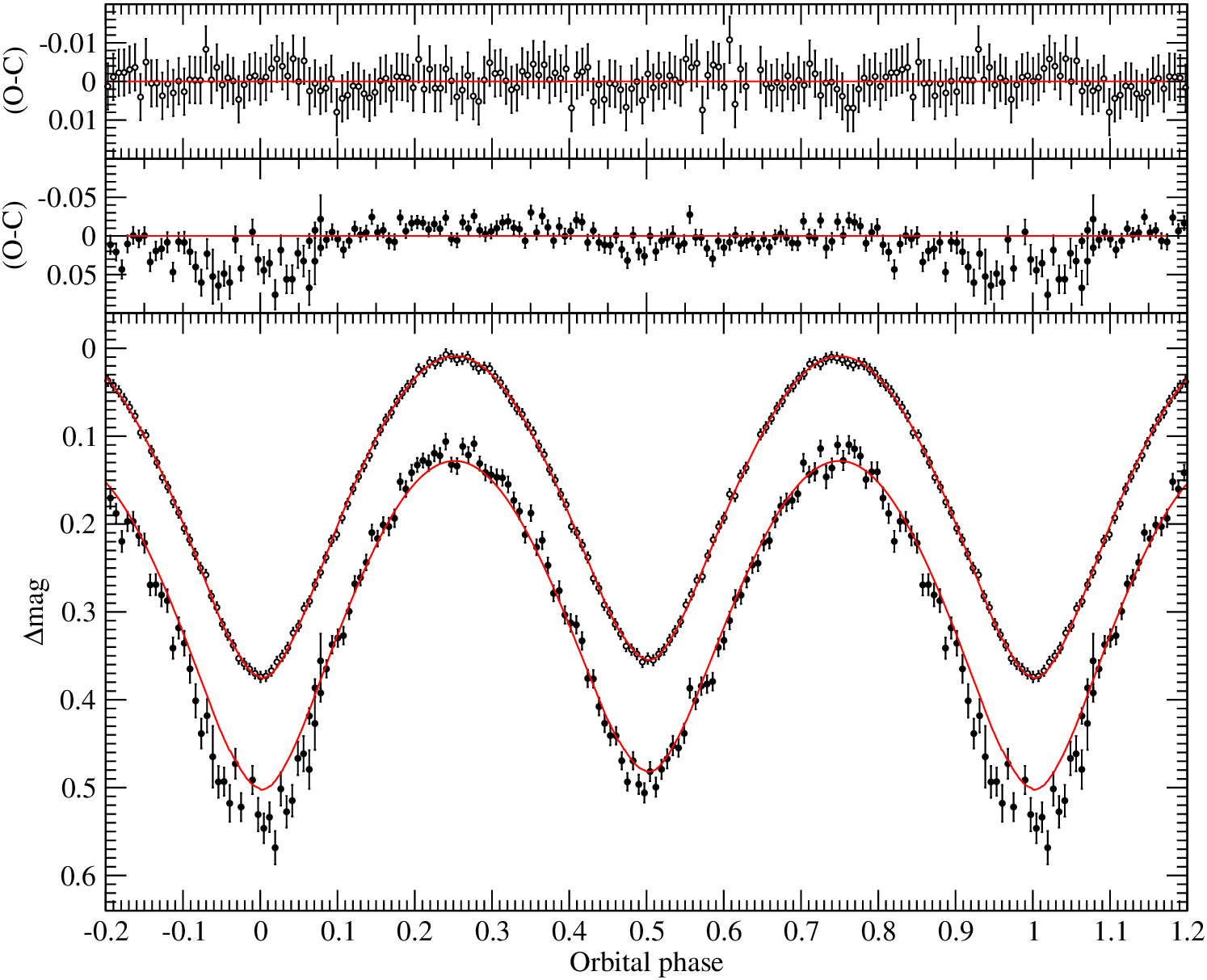}
   \caption{Light  curve fits  assuming an  overcontact binary  system
     made of two degenerate post-AGB cores. B- and I-band light curves
     are  shown in  the lower  panel  using filled  and open  circles,
     respectively.  Red  lines  are   the  best-fit  models  for  each
     band. Residuals are shown in the upper panels. Note the different
     scale for each band.}
   \label{fig:overcontact}
   \end{center}
\end{figure}

\begin{table}
\begin{center}
\begin{tabular}{lcc}
\hline
\hline
 & Primary & Secondary \\
\hline
$M$ ($M_{\sun}$)          & 0.48 ($0.22-1.50$)    & 0.47 ($0.22-1.48$)  \\
$R$ ($R_{\sun}$)          & 0.57 ($0.44-0.84$)    & 0.57 ($0.44-0.83$)  \\
\hline
$r$ pole                 & 0.400 ($0.386-0.420$) & 0.397 ($0.384-0.415$)\\
$r$ point                & contact               & contact              \\     
$r$ side                 & 0.431 ($0.415-0.453$) & 0.426 ($0.412-0.448$)\\
$r$ back                 & 0.495 ($0.486-0.515$) & 0.490 ($0.486-0.510$)\\
\hline
\end{tabular}
\end{center}
\caption{Absolute masses  and volumetric  radii derived from  the best
  fit to light curves assuming  an overcontact binary system. For each
  component, the  radius relative  to the semimajor  axis is  given in
  different directions  accounting for  deformation: towards  the pole
  ($r_{\rm  pole}$)  and  towards  the  companion  direction  ($r_{\rm
  point}$),  opposite  ($r_{\rm  back}$),  and  perpendicular  to  the
  orbital plane ($r_{\rm side}$).  Numbers in parentheses indicate the
  range of possible values corresponding  to fits with $0.8\leq q \leq
  1.2$ and $1.0\leq a \leq 1.9\, R_{\sun}$.
  \label{tab:overcontact}}
\end{table}

Given that  to fit the  observational data the secondary  component in
the case of semidetached system  should have a luminosity nearly equal
to that of  the primary we propose an alternative  model in which both
stars  are at  the same  evolutionary  stage.  However,  we relax  the
assumptions  of \cite{SantanderGarciaetal2015}  to  allow  for a  more
physically  motivated solution.   In  particular, we  assume that  the
position   of  Henize~2--428   in   the  Hertzsprung-Russell   diagram
corresponds to  that of a  model post-AGB  star of mass  $\simeq 0.5\,
M_{\sun}$, red line in Fig.~\ref{fig:HR}. Thus, we adopt again $T_{\rm
eff,1}=45,000$~K. We note that since this mass is very similar to that
found when  a semi-detached system  is considered we keep  the orbital
separation at $a=1.27\, R_{\sun}$.  Additionally  we did not take into
account the  mass ratio  and the semi-major  axis they  derive because
they are mainly constrained by the radial velocity fits, which we have
argued  that  may  not  be  related   to  the  binary  nature  of  the
system. Nevertheless,  the distance to  the object and  its luminosity
may provide constraints  on the scale of the  binary system. Following
our  previous estimates  in the  case of  a semi-detached  binary, the
total luminosity  of the  system is  scaled by  $L= 845  (D/1.4\, {\rm
kpc})^{2}\,  L_{\sun}$.   If  we  consider  the  independent  distance
measurements    listed    earlier    in    Sect.~\ref{subsec:stellar},
Henize~2--428 may  be at distance between  1.7~kpc and $2.7\pm0.5$~kpc
\citep{Maciel1984,Frew}.   Then, the  total luminosity  of the  system
would be between 1,246 and  4,415~$L_{\sun}$.  We did not consider the
distance   estimate  of   \citep{SantanderGarciaetal2015},  which   is
shorter, to keep our estimates independent  of the light curve data we
are using.  Complementary,  the total luminosity can  also be computed
from the effective  temperatures, and relative radii  derived from the
light curve fits, along with  the assumed semi-major axis.  Therefore,
the distance limits may provide  constraints on the orbital semi-major
axis.

Taking  all this  into account  we have  performed fits  to the  light
curves assuming different  values of mass ratio (from 0.5  to 1.5) and
semi-major  axis  (from  1.0  to  $2.0\,  R_{\sun}$),  fixing  $T_{\rm
eff,1}=45,000$~K.  The resulting  light curves for our  best fit model
are shown  in Fig.~\ref{fig:overcontact}, and the  physical parameters
of the  binary system  are listed in  Table~\ref{tab:overcontact}.  We
note that, as expected, neither the semi-major axis nor the mass ratio
have a significant impact  on the quality of the fits  at the level of
precision of the light curves. However, they have an obvious impact on
the absolute properties  of the system. On the one  hand, the absolute
radius of  each component  is proportional to  $a$, which  defines the
scale of  the system.  On  the other hand,  the relative size  of each
component ($r$, derived from  the pseudo-potential) is correlated with
the mass  ratio because the stars  of the binary are  in contact. From
the relative radii, the effective temperature and the semi-major axis,
we computed the total luminosity of  the system for each fit ($L\simeq
( a\,  r )^2 T_{\rm  eff}^4$).  The  results show that  the previously
computed  luminosity  limits are  consistent  with  a semi-major  axis
ranging  from 1.0  to $1.9\,  R_{\sun}$, almost  independently of  the
value of the mass ratio.  Given the  period of the system, this can be
translated  into a  total  mass of  the system  in  the range  $\simeq
0.45-3.0\,  M_{\sun}$.  If  we a  assume a  mean measured  distance of
2.2~kpc, then  $a \simeq 1.3\, R_{\sun}$  and the total mass  would be
$\simeq 1.0\, M_{\sun}$, in agreement  with the position of the system
on the Hertzsprung-Russell diagram.

Given that  the relative  size of the  components correlates  with the
mass  ratio  of the  system,  we  have  run  fits by  fixing  $a=1.3\,
R_{\sun}$ and  setting the mass ratio  as a free parameter.   The best
fit is found  for $q=0.98\pm0.20$, very close to the  ratio adopted by
\cite{SantanderGarciaetal2015}, although  it is poorly  constrained by
the light curves.   As can be seen  in Fig.~\ref{fig:overcontact}, for
this model the agreement between  the theoretical light curves and the
observed ones  is excellent,  both in  the B-band  and in  the I-band,
being   the   residuals   very  small,   16.93~mmag   and   3.25~mmag,
respectively.  For the  best fit model, the inclination  of the system
is $i=63.2^\circ$ and the effective  temperature of the secondary star
is $T_{\rm eff,2}=41,100$~K. These parameters yield the masses of both
components  of  the  binary  system, $0.48\,  M_{\sun}$,  and  $0.47\,
M_{\sun}$.   Thus, the  total mass  of  the binary  system is  $0.95\,
M_{\sun}$, well below the Chandrasekhar  limiting mass. Hence, even if
the system is  made of two post-AGB stars, the  total combined mass of
the system will not be enough to produce a Type Ia supernova outburst,
as \cite{SantanderGarciaetal2015}  claimed.  In  any case,  stars with
these properties are  most probably post-AGB stars,  and the component
of lower mass is likely to be a helium white dwarf.

Nevertheless,  we stress  here  that the  absolute  properties of  the
components of Henize~2--428 are only  roughly constrained by the light
curve  fits if  we take  into  account the  range of  mass ratios  and
orbital  semi-major axes  that produce  comparable fits  to the  light
curves.  Table~\ref{tab:overcontact} shows the  properties of the best
fit described above,  with the range of possible  values associated to
the mass ratio and semimajor axis uncertainties. Although the combined
analysis of light curves and  the position in the Hertzsprung--Russell
diagram  favor a  scenario of  two similar  stars with  masses $\simeq
0.5\, M_{\sun}$,  only precise radial velocities  can better constrain
the model.

\section{Summary and conclusions}
\label{sec:summary}

We addressed  the recent claim made  by \cite{SantanderGarciaetal2015}
that the central  binary system of the  planetary nebula Henize~2--428
is  a  SN~Ia   progenitor  in  the  frame   of  the  double-degenerate
scenario.  This claim  has attracted  some attention,  and if  true it
would be the first super-Chandrasekhar  mass binary white dwarf system
with short  period discovered ever.   More interestingly, it  would be
definitely  located in  the center  of a  planetary nebula,  providing
strong  evidence   for  SN~Ia  occurring  inside   PNe,  termed  SNIPs
\citep{TsebrenkoSoker2015}.    However,   we  found   that,   although
certainly possible, this claim is  not yet univocally supported by the
observations, and  that different explanations are  possible, and need
to be considered.

Our statement is  based not only on  purely theoretical considerations
about   the  evolutionary   properties  of   the  central   object  of
Henize~2--428, but  also on  the way  the observed  data set  has been
analyzed.  In particular,  in Sect.~\ref{sec:considerations} we argued
that  the claim  for an  equal-masses binary  system and  the required
mass, luminosity,  and radius of  the two  stars does not  comply with
evolutionary  tracks of  post-AGB stars  (see Fig.~\ref{fig:HR}).   We
also  analyzed   critically  the   way  in  which   observations  were
interpreted, and we questioned the two most important assumptions made
to study the observed properties of the binary system at the center of
Henize~2--428.  These  are that both  components of the  binary system
have exactly the  same mass, and that, moreover, they  are at the same
evolutionary stage.  

In particular, in Sect.~\ref{subsec:stellar}  we argued that there are
strong  theoretical arguments  that  pose a  problem  to the  scenario
proposed by \cite{SantanderGarciaetal2015}. One  of these arguments is
that that  the initial mass  difference in  the main sequence  must be
very small.  The second one is that the population of twin binaries is
small. Furthermore,  in Sect.~\ref{subsec:observations}  we critically
examined the  joint analysis of the  light curves and the  spectrum of
\cite{SantanderGarciaetal2015}.   Specifically,  we  argued  that  the
explanation  of \cite{SantanderGarciaetal2015}  of the  variability of
the  He~{\sc II}~5412~\AA~absorption  spectral  line  as arising  from
Doppler shifts of two absorption lines, one from each star, is not the
only possible one.  Instead, we  suggested that the variability of the
He~{\sc II}~5412~\AA~spectral line can be accounted for by a, possibly
time-varying, broad  absorption line from  the central star on  top of
which there  is a time-varying  narrow emission line from  the compact
nebula or even from much closer to the star.  \cite{Dobrincicetal2008}
find the age  of the Henize~2-428 equatorial ring to  be $4,300 (D/1.8
\kpc) \yr$.  Although the ring is old, there is a compact dense nebula
near the central star \citep{Rodriguezetal2001}.  It is quite possible
that the He~{\sc II}~5412~\AA~narrow emission spectral line sitting on
top of the broad absorption line  originates in the compact nebula, or
from an outflow that feeds the compact nebula much closer to the star.
This explanation could also be  compatible with the observed spectrum,
but  more  detailed studies  and  better  observations are  needed  to
resolve in detail the spectral feature.

All these  considerations led us to  judge that although the  first of
the assumptions of \cite{SantanderGarciaetal2015} -- namely, that both
stars  have approximately  the same  mass  -- is  quite possible,  the
second one -- that both components of the binary system are at exactly
the same  evolutionary stage -- is  not fully justified, and  needs an
independent  evaluation.   Since  these   two  assumptions  are  quite
restrictive, we explored alternative  models. Accordingly, we analyzed
other  possible explanations,  less extraordinary  but that  could fit
equally well  the observations and, simultaneously,  do not contradict
stellar evolutionary results (Sect.~\ref{sec:model}).

The first  of these  scenarios still  hold the  existence of  a binary
system, as the two (almost) symmetric  broad minima in the light curve
are attributed  to tidal distortion  caused by a companion  of similar
mass.  In particular, we mentioned  a binary system composed of either
a post-RGB or a post-AGB star with a low-mass companion.  In the first
case, a low-mass main sequence star  truncates the evolution of a star
of initial mass $M_{{\rm ZAMS}_1} \approx 1.0\, M_{\sun}$ on the upper
RGB.   In the  second scenario,  a  main sequence  star truncates  the
evolution of a star of  initial mass $M_{{\rm ZAMS}_1} \approx 2.5-3.0
M_{\sun}$ on the lower AGB. However, we found that although this model
is  plausible  it  does  not  fit  well  the  observed  light  curves.
Consequently, we studied a second possibility in which the assumptions
of \cite{SantanderGarciaetal2015} are relaxed.  We found that a binary
system made  of two  post-AGB stars of  masses $0.453\,  M_{\sun}$ and
$0.437\, M_{\sun}$  can fit equally  well the observed light  curve of
the  system. Hence,  if our  interpretation  of the  observed data  is
correct  the  combined  mass  of  the binary  system  would  be  below
Chandrasekhar's limiting  mass, and it  would not explode as  a SN~Ia.
We conclude this  short study by stating that the  exact nature of the
central object of Henize~2--428 is still to be determined.

\section*{Acknowledgements}
We acknowledge  useful discussions  with Miguel  Santander-Garc\'\i a,
who also  supplied us unpublished  data.  This research  was partially
supported by MINECO grant AYA2014-59084-P, by the AGAUR (EG-B), by the
AGENCIA  through the  Programa  de  Modernizaci\'on Tecnol\'ogica  BID
1728/OCAR, and by  the PIP 112-200801-00940 grant  from CONICET (LGA).
IR  acknowledges support  from the  MINECO  and the  Fondo Europeo  de
Desarrollo  Regional (FEDER)  through grants  ESP2013-48391-C4-1-R and
ESP2014-57495-C2-2-R.

\bibliographystyle{elsarticle-harv}

\end{document}